\documentclass[12pt]{article}
\usepackage{authblk}
\usepackage{graphicx}
\usepackage{float}
\usepackage{subcaption}
\usepackage[a4paper,margin=1in]{geometry}
\usepackage{booktabs}
\usepackage{caption}
\usepackage{adjustbox}
\graphicspath{{figures/}}
\usepackage{hyperref}
\usepackage{amsmath}
\usepackage{mathrsfs} 
\usepackage{titlesec}
\usepackage{forest}
\usepackage{dingbat}

\usepackage[margin=1in]{geometry}

\usepackage{setspace} 
\title{\textbf{From Claims to Evidence: A Unified Framework and Critical Analysis of CNN vs. Transformer vs. Mamba in  Medical Image Segmentation}}

\author{
\fontsize{11pt}{13pt}\selectfont 
Pooya Mohammadi Kazaj, MSc\textsuperscript{1}; Giovanni Baj, PhD\textsuperscript{1}; Yazdan Salimi, PhD\textsuperscript{2}; Anselm W. Stark, MD\textsuperscript{1}; Waldo Valenzuela, PhD\textsuperscript{3}; George CM. Siontis, MD, PhD\textsuperscript{1}; Habib Zaidi, PhD\textsuperscript{2,4,6};
Mauricio Reyes, PhD\textsuperscript{6,7,8}; Christoph Gräni, MD, PhD\textsuperscript{1}; Isaac Shiri, PhD\textsuperscript{1} \\
\begin{flushleft}
\textsuperscript{1}Department of Cardiology, Inselspital, Bern University Hospital, University of Bern, Bern, Switzerland \\
\textsuperscript{2}Geneva University Hospital, Division of Nuclear Medicine and Molecular Imaging, CH-1211 Geneva, Switzerland \\
\textsuperscript{3}University Institute for Diagnostic and Interventional Neuroradiology, Inselspital, Bern University Hospital, University of Bern, Freiburgstrasse, 3010, Bern, Switzerland \\
\textsuperscript{4}Department of Nuclear Medicine and Molecular Imaging, University of Groningen, University Medical Center Groningen, Groningen, Netherlands \\
\textsuperscript{5}Department of Nuclear Medicine, University of Southern Denmark, Odense, Denmark \\
\textsuperscript{6}Department of Radiation Oncology, University Hospital Bern, University of Bern, Bern, Switzerland \\
\textsuperscript{7}Center for Artificial Intelligence in Medicine, University of Bern, Bern, Switzerland \\
\textsuperscript{8}ARTORG Center for Biomedical Engineering Research, University of Bern, Bern, Switzerland \\
\end{flushleft}
}

\titleformat{\section}{\normalfont\Large\bfseries}{\thesection}{1em}{}
\titleformat{\subsection}{\normalfont\large\bfseries}{\thesubsection}{1em}{}
\titleformat{\subsubsection}{\normalfont\normalsize\bfseries}{\thesubsubsection}{1em}{}
\titleformat{\paragraph}[runin]{\normalfont\normalsize\bfseries}{\theparagraph}{1em}{} 
\date{}

\begin{document}
\maketitle

\vspace{1em}
\noindent \textbf{Corresponding Author:} \\
Isaac Shiri, Ph.D.\\
Department of Cardiology \\
University Hospital Bern \\
University of Bern\\
Freiburgstrasse \\
CH - 3010 Bern, Switzerland \\
Email: isaac.shirilord@unibe.ch \\

\newpage
\begin{abstract}
While numerous architectures for medical image segmentation have been proposed, achieving competitive performance with state-of-the-art models networks such as nnUNet, still leave room for further innovation.  In this work, we introduce nnUZoo, an open source benchmarking framework built upon nnUNet, which incorporates various deep learning architectures, including CNNs, Transformers, and Mamba-based models. Using this framework, we provide a fair comparison to demystify performance claims across different medical image segmentation tasks. Additionally, in an effort to enrich the benchmarking, we explored five new architectures based on Mamba and Transformers, collectively named X$^2$Net, and integrated them into nnUZoo for further evaluation. The proposed models combine the features of conventional U$^2$Net, nnUNet, CNN, Transformer, and Mamba layers and architectures, called X$^2$Net (UNETR$^2$Net (UNETR), SwT$^2$Net (SwinTransformer), SS2D$^2$Net (SwinUMamba),  Alt1DM$^2$Net (LightUMamba), and  MambaND$^2$Net (MambaND)). We extensively evaluate the performance of different models on six diverse medical image segmentation datasets, including microscopy, ultrasound, CT, MRI, and PET, covering various body parts, organs, and labels. We compare their performance, in terms of dice score and computational efficiency, against their baseline models, U$^2$Net, and nnUNet. CNN models like nnUNet and U$^2$Net demonstrated both speed and accuracy, making them effective choices for medical image segmentation tasks. Transformer-based models, while promising for certain imaging modalities, exhibited high computational costs. Proposed Mamba-based X$^2$Net architecture (SS2D$^2$Net) achieved competitive accuracy with no significantly difference from nnUNet and U$^2$Net, while using fewer parameters. However, they required significantly longer training time, highlighting a trade-off between model efficiency and computational cost. All developed models are available on: \href{https://github.com/AI-in-Cardiovascular-Medicine/nnUZoo}{https://github.com/AI-in-Cardiovascular-Medicine/nnUZoo}.

\textbf{Keywords:} Deep Learning; Segmentation; Benchmark; CNN; Mamba; Transformer

\end{abstract}
\newpage

\section{Introduction}
Medical image segmentation is essential for computer-assisted diagnosis, quantitative image analysis, and treatment planning in various imaging modalities and clinical applications. Although numerous deep learning-based architectures for medical image segmentation have been proposed, achieving competitive performance with state-of-the-art models like nnUNet~\cite{isensee_nnu-net_2021} and Auto3dseg~\cite{Cardoso_MONAI_An_open-source_2022}, still leaves room for further innovation. Despite their success, there is a need for new architectures that can drive further improvements in performance, robustness, generalization, adaptability, and efficiency \cite{gut2022benchmarking,Ise_nnUNet_MICCAI2024}.
In addition to architectural advances of deep neural networks, the development of medical image segmentation networks relies on various tools and libraries. However, many of these frameworks are designed for general computer vision tasks rather than being specialized for medical imaging. Unlike natural images, each medical imaging modality and task requires its own preprocessing steps adapted to its specific characteristics to ensure compatibility with deep learning models. On the other hand, libraries such as MONAI~\cite{Cardoso_MONAI_An_open-source_2022} have been specifically developed for deep learning-based medical image segmentation, providing domain-specific tools and workflows. Additionally, nnUNet remains a state-of-the-art framework that has demonstrated remarkable adaptability across different segmentation tasks in different imaging modalities and tasks. A recent revision of nnUNet~\cite{Ise_nnUNet_MICCAI2024} has shown that many newly proposed architectures, which claim superior performance, do not exceed nnUNet when evaluated under fair comparison scenarios. This presents an opportunity to use the robust preprocessing pipeline of nnUNet as a foundation for the development of novel architectures, further improving segmentation performance in medical imaging.

Convolutional neural network (CNN)-based architectures, particularly UNet ~\cite{U-Net} and its variants, have long been the backbone of different image segmentation tasks. Recently, Transformer-based~\cite{vaswani2017attention} architecture, such as UNETR~\cite{hatamizadeh_unetr_2022} as well as state-space models like Mamba~\cite{mamba}, have emerged as promising architectures. CNNs efficiently capture spatial hierarchies with local receptive fields but struggle with long-range dependencies. Transformers use self-attention to address this but are computationally demanding. Their performance difference mainly stems from inductive bias—Transformers, being more generic, require larger datasets and perform well in computer vision when trained with sufficient samples. Mamba, a state-space model (SSM), scales linearly with sequence length, making it an efficient alternative to Transformers ~\cite{mamba,Xin_SegMamba_MICCAI2024}. While CNNs and Transformers remain dominant in medical imaging, SSMs introduce advantages in dependency modeling, although challenges in efficiency and adaptability persist~\cite{mamba,Xin_SegMamba_MICCAI2024}.

Beyond architectural innovations (CNN, Transformer, and Mamba), increasing network depth introduces additional challenges, particularly in maintaining a balance between computational cost and efficiency. Various approaches, such as stacking multiple UNet models to form cascade networks, have been proposed to address this; however, these methods scale memory and computation significantly depending on the number of stacked models. U$^2$Net~\cite{Qin_2020_PR} addresses this issue by introducing a nested U-structure, enabling deeper architectures while maintaining efficiency and minimizing additional resource demands. This design allows for improved feature extraction without the exponential increase in computational burden seen in cascade models.

Despite these advances, the question remains: Do these increasingly complex architectures provide a tangible improvement over well-established models like nnUNet, or do they merely shift computational trade-offs without delivering significant benefit? A systematic benchmarking effort is required to fairly assess these new models under controlled conditions, ensuring that the observed performance gains are not confounded by differences in training protocols, data preprocessing, or hyperparameter tuning.
In this work, we introduce nnUZoo, an open-source library and benchmarking framework built upon nnUNet, leveraging its robust preprocessing pipeline. nnUZoo incorporates various deep learning architectures, including CNNs, Transformers, and Mamba-based models, to provide a fair comparison and demystify performance claims across different medical image segmentation tasks in low code mode approach. Furthermore, in an effort to enrich the benchmarking, we explored five new architectures based on Mamba and Transformers, collectively named X$^2$Net, and integrated them into nnUZoo for further evaluation.

\newpage
\section{Methods}

\subsection{Datasets and Evaluation Metrics}

We used six diverse datasets from different medical image modalities (microscopy~\cite{microscopy}, ultrasound~\cite{camus}, CT~\cite{abdomen_ct}, MRI~\cite{abdomen_mri,acdc}, and PET~\cite{pet_data}) covering various body parts, organs, and labels to evaluate the developed models and pipeline. The dataset selection process follows the approach used in ~\cite{umamba}. Details of the datasets are provided in Table~\ref{data-info}. Furthermore, models were evaluated with the dice score and the performance was compared using the Wilcoxon sigend-rank~\cite{wilcoxon} test.

\subsection{Model Zoo}

\subsubsection{Benchmarking of existing Model in nnUZoo:}
In addition to the nnUNet pipeline(version 2.1.1), we implemented various architectures, including CNN-based models (U$^2$Net, and U$^2$Net small), a Transformer-based model (UNETR and SwinTransformer~\cite{liu_swt_2021} (SwT)), and a Mamba-based model (SwinUMamba~\cite{Swin-UMamba}, LightUMamba~\cite{liao2024lightmunet}, and SegMamba~\cite{Xin_SegMamba_MICCAI2024}. These baseline models were used with no modifications or very few, compared to their original implementations. 

\subsubsection{Implementing X$^2$Net in nnUZoo:}
\begin{figure}
    \centering
    \includegraphics[width=\textwidth]{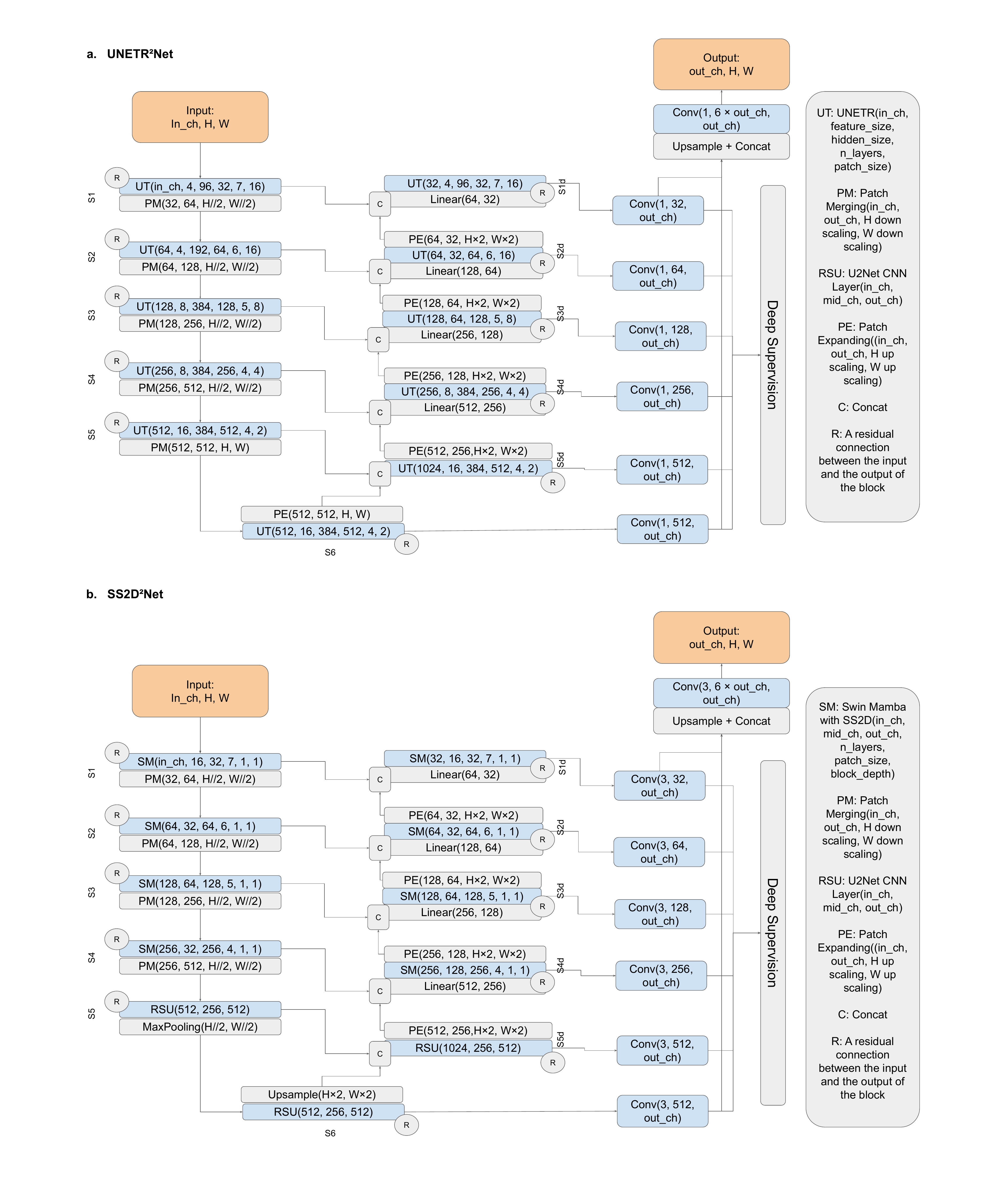}
    \caption{\textit{X$^2$Net architectures implemented in this study. \textbf{a.} SS2D$^2$Net. b. UNETR$^2$Net. SwT$^2$Net and Alt1DM$^2$Net follow the same structure of SS2D$^2$Net in which SM is changed with SwinTransformer and LightUMamba, respectively. Likewise, MambaND$^2$Net follows the same architecture of UNETR$^2$Net in which UT is changed with MambaND U-Block. These variants are proposed to enrich nnUZoo to merge between beneficial properties of previously proposed models}}
    
\label{model-graphs}
\end{figure}

The proposed UNETR$^2$Net model, based on the U$^2$Net architecture, is shown in Figure~\ref{model-graphs} (a). To reduce the number of parameters in UNETR, we decreased the feature size and embedding dimensions. Additionally, each UNETR U-Block was modified to accommodate varying input patch sizes, as a result, merging and expansion layers are done automatically with varying scale sizes. We refer to each Block in U$^2$Net-like architecture as U-Block. Similar to residual blocks, the input of each UNETR U-Block is added to its output, adapted through a depth-wise separable convolution layer~\cite{howard_mobilenets_2017} . We also propose SwT$^2$Net, which leverages Swin Transformers as U-Blocks and benefits from residual depth-wise separable connections and ReSidual U-Blocks(RSU) introduced in U$^2$Net.

The proposed SSD$^2$Net model, based on the U$^2$Net architecture, is shown in Figure~\ref{model-graphs} (b). This model combines SwinUMamba and U$^2$Net, integrating four layers of SwinUMamba U-Blocks, continued by three ReSidual U-Blocks (RSU), and followed by four SwinUMamba U-Blocks as the decoder. SwinUMamba layers utilize 2D Selective Scan (SS2D)~\cite{liu2024vmamba} module. SS2D is a modified Mamba module that traverses the input data in four directions instead of the single direction used in the primary Mamba layers. Like all U$^2$Net models, both the CNN and Mamba blocks incorporate residual connections, where the input is added to the output using depthwise separable convolution layers. Moreover, similar to UNETR, the Mamba layers are modified to accept different input sizes by varying the merging and patching layers. In conclusion, this design aims to leverage the strengths of CNNs and state-space models to achieve a balance between efficiency and accuracy. SS2D$^2$NetS on the other hand, is a more compact version of SS2D$^2$Net, designed to reduce computational costs while maintaining competitive segmentation performance. 

\begin{table}

\caption{Dataset Descriptions of different medical image modalities used for training and evaluation of different models.}
\label{data-info}
\resizebox{\linewidth}{!}{%
\begin{tabular}{|c|c|c|c|c|c|}
\hline
Tasks & Train images & Test images & Classes & Patch Size  & Batch Size\\
\hline
Microscopy~\cite{microscopy} & 1000 & 101 & 2 & 256 x 256  & 8\\
CAMUS~\cite{camus} & 900 & 100 & 3 & 256 x 256  & 6 \\
ACDC~\cite{acdc} & 200 (1902 slices)  & 100      (1076 slices) & 4 & 256 x 224  & 16\\
AbdomenMR~\cite{abdomen_mri} &  60 (5615 slices)  & 50 (3357 slices) & 14 & 320 x 320  & 8\\
AbdomenCT~\cite{abdomen_ct} &  50 (4794 slices) & 50 (10894 slices) & 14 & 256 x 256  & 6\\
PET~\cite{pet_data} & 50  (29606 slices) & 12 (9074 slices) & 23 & 320 x 192  & 4\\
\hline

\end{tabular}
}
\end{table}
In the same spirit, we employed LightUMamba as the U-Block and incorporated the alternating traversing mechanism introduced in MambaND~\cite{li_mamba-nd_2024} in which for each block the traversing mechanism changes and utilized Gated Spatial Convolution (GSC) to integrate local spatial feature extraction of CNNs to Mamba layers, as proposed in SegMamba~\cite{Xin_SegMamba_MICCAI2024}. Building on this, we propose Alt1DM$^2$Net and Alt1DM$^2$NetS, which are lighter models compared to SS2D$^2$Net. We also integrated MambaND~\cite{li_mamba-nd_2024} into the X$^2$Net architecture, which is called MambaND$^2$Net. This architecture is fully based on MambaND U-Blocks with no RSU layers similar to UNETR$^2$Net in Figure~\ref{model-graphs} (a).

\subsection{Training Setup}

\begin{table}

\centering
\caption{Dice scores of different models on test datasets. The results vary across different datasets and models, with the highest-performing model highlighted in bold for each dataset.}
\label{dice-table}
\resizebox{\linewidth}{!}{%
\begin{tabular}{|c|c|c|c|c|c|c|}
\toprule
Models &                      Microscopy &                           CAMUS &                            ACDC &                              AbdomenMR &                              AbdomenCT &                             PET \\
\midrule
nnUNet                      &                0.69 \textpm 0.20 &  \textbf{0.92 \textpm 0.03} &  \textbf{0.92 \textpm 0.03} &  \textbf{0.74 \textpm 0.11} &                0.78 \textpm 0.08 &  \textbf{0.73 \textpm 0.04} \\
UNETR                       &  \textbf{0.72 \textpm 0.23} &                0.89 \textpm 0.04 &                0.89 \textpm 0.04 &                0.56 \textpm 0.18 &                0.47 \textpm 0.20 &                0.50 \textpm 0.03 \\
SwT                         &                0.66 \textpm 0.21 &                0.89 \textpm 0.04 &                0.91 \textpm 0.03 &                0.61 \textpm 0.14 &                0.60 \textpm 0.13 &                0.50 \textpm 0.03 \\
SwinUMamba                    &                0.70 \textpm 0.21 &  \textbf{0.92 \textpm 0.03} &                0.90 \textpm 0.03 &                0.73 \textpm 0.13 &                0.78 \textpm 0.09 &                0.71 \textpm 0.04 \\
SegMamba                    &                0.71 \textpm 0.20 &                0.91 \textpm 0.03 &  \textbf{0.92 \textpm 0.03} &                0.70 \textpm 0.17 &                0.75 \textpm 0.11 &                0.72 \textpm 0.04 \\
LightUMamba                     &                0.70 \textpm 0.22 &                0.91 \textpm 0.03 &  \textbf{0.92 \textpm 0.02} &                0.71 \textpm 0.15 &                0.73 \textpm 0.11 &                0.71 \textpm 0.04 \\
\midrule
U\textsuperscript{2}Net     &                0.70 \textpm 0.21 &  \textbf{0.92 \textpm 0.03} &  \textbf{0.92 \textpm 0.03} &                0.73 \textpm 0.15 &                0.78 \textpm 0.08 &                0.72 \textpm 0.04 \\
U\textsuperscript{2}NetS    &                0.69 \textpm 0.22 &                0.91 \textpm 0.04 &  \textbf{0.92 \textpm 0.03} &                0.72 \textpm 0.13 &                0.71 \textpm 0.10 &                0.65 \textpm 0.04 \\
\midrule
UNETR\textsuperscript{2}Net   &                0.68 \textpm 0.23 &                0.87 \textpm 0.05 &                0.91 \textpm 0.03 &                0.65 \textpm 0.15 &                0.69 \textpm 0.13 &                0.66 \textpm 0.04 \\
SwT\textsuperscript{2}Net   &                0.70 \textpm 0.19 &                0.91 \textpm 0.03 &                0.91 \textpm 0.03 &                0.65 \textpm 0.15 &                0.71 \textpm 0.10 &                0.67 \textpm 0.04 \\
SS2D\textsuperscript{2}Net  &                0.71 \textpm 0.19 &  \textbf{0.92 \textpm 0.03} &  \textbf{0.92 \textpm 0.03} &  \textbf{0.74 \textpm 0.13} &  \textbf{0.80 \textpm 0.08} &                0.72 \textpm 0.04 \\
SS2D\textsuperscript{2}NetS &                0.69 \textpm 0.20 &                0.90 \textpm 0.04 &  \textbf{0.92 \textpm 0.03} &                0.69 \textpm 0.15 &                0.72 \textpm 0.11 &                0.64 \textpm 0.04 \\
Alt1DM\textsuperscript{2}Net  &                0.67 \textpm 0.23 &                0.91 \textpm 0.04 &                0.91 \textpm 0.03 &                0.65 \textpm 0.17 &                0.74 \textpm 0.09 &                0.68 \textpm 0.04 \\
Alt1DM\textsuperscript{2}NetS &                0.70 \textpm 0.19 &                0.89 \textpm 0.04 &                0.91 \textpm 0.04 &                0.64 \textpm 0.16 &                0.63 \textpm 0.13 &                0.57 \textpm 0.03 \\
MambaND\textsuperscript{2}Net   &                0.66 \textpm 0.21 &                0.91 \textpm 0.04 &                0.89 \textpm 0.04 &                0.64 \textpm 0.14 &                0.69 \textpm 0.11 &                0.67 \textpm 0.05 \\
\bottomrule
\end{tabular}
}
\end{table}

All models were integrated into the nnUNet architecture, serving as the foundational framework. The integration was performed using a modified version of nnUNet, referred to as nnUzoo. All models were trained for 250 epochs, utilizing an 80/20\%  (train/validation)  split of the training dataset for a single fold. The training and validation datasets were kept the same for all the training setups.  Moreover, the preprocessing, loss function which was a weighted sum of DiceLoss and CrossEntroyLoss, and training augmentations were also kept the same for all the models from nnUNet. To ensure uniformity and compatibility with hardware constraints, key hyperparameters such as batch size and patch size were standardized across all models, as detailed in Table~\ref{data-info}. These adjustments were specifically designed to accommodate the 40 GB of VRAM available, particularly addressing the higher memory demands of non-CNN-based models. This approach ensured a fair comparison and efficient utilization of computational resources during the training process, while maintaining consistency in experimental conditions across all models. All developed models are available on \href{https://github.com/AI-in-Cardiovascular-Medicine/nnUZoo}{https://github.com/AI-in-Cardiovascular-Medicine/nnUZoo}.
\begin{figure}
    \centering
    \includegraphics[width=\textwidth]{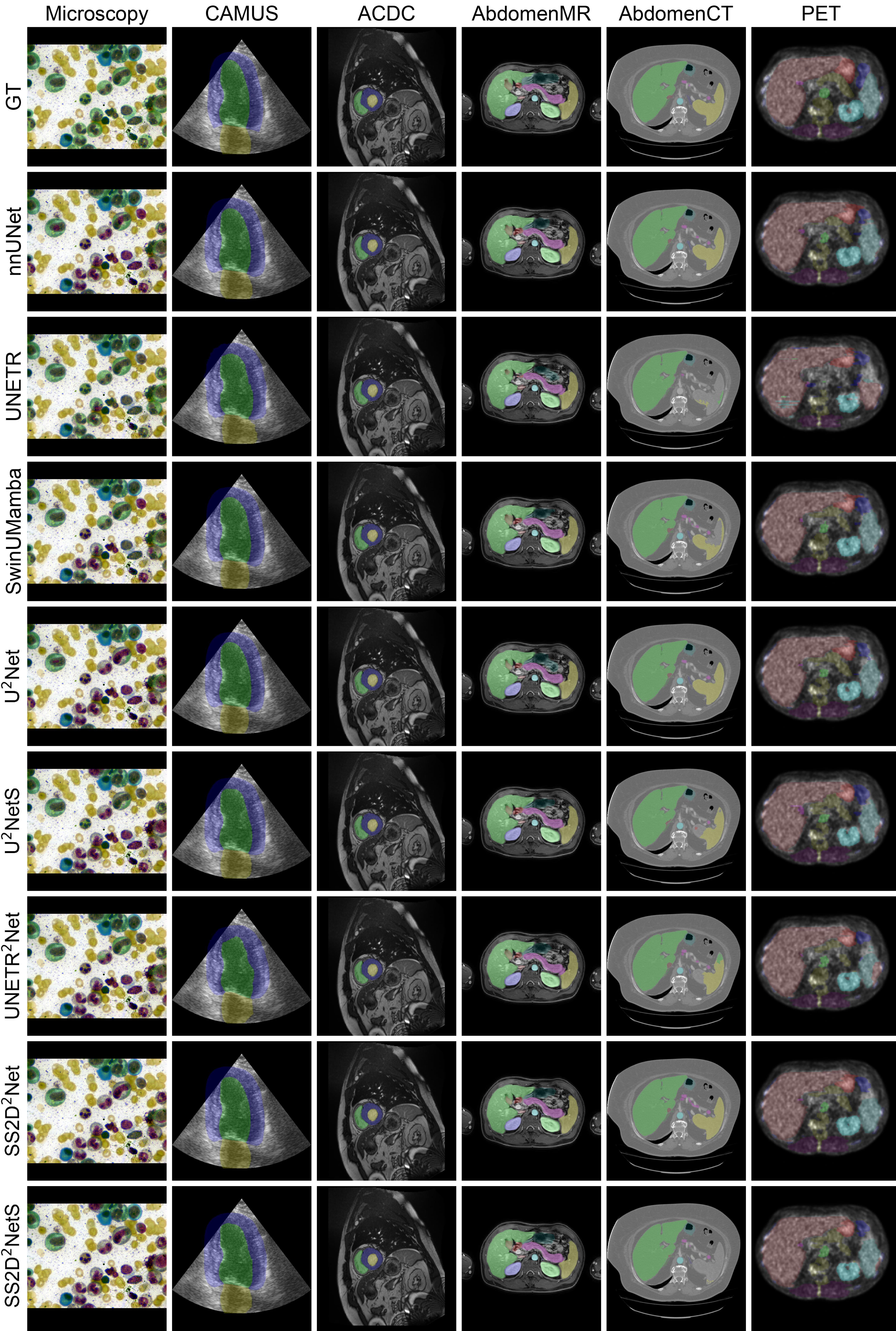}
    \caption{\textit{Example results of the best models on all datasets. GT (first row) corresponds to the ground truth segmentation.}}
    \label{fig:dice-comparison}
\end{figure}

\newpage
\section{Results and Discussion}
Figure~\ref{fig:dice-comparison} presents examples of segmentation outputs from best models across various imaging modalities, highlighting the variation in network performance depending on the modality.

The quantitative performance of different networks across various datasets is summarized in Table~\ref{dice-table} in terms of the dice score, which represents the average dice score across all labels in different imaging modalities. As highlighted in Table~\ref{dice-table}, performance differences are modality-dependent. For instance, across microscopy, CAMUS, and ACDC datasets, all networks performed almost similarly with slight differences. However, for more complex structures and labels, such as in AbdomenMR, AbdomenCT, and PET datasets, the UNETR and SwinTransformer baselines and their X$^2$Net variants (UNETR$^2$Net and SwT$^2$Net) did not perform well despite having the largest number of parameters, showing significantly lower performance compared to other networks. SwinUMamba, LightUMamba, and SegMamba, however, performed comparably to nnUNet across most modalities.
\begin{figure}

    \centering
    \includegraphics[width=\textwidth]{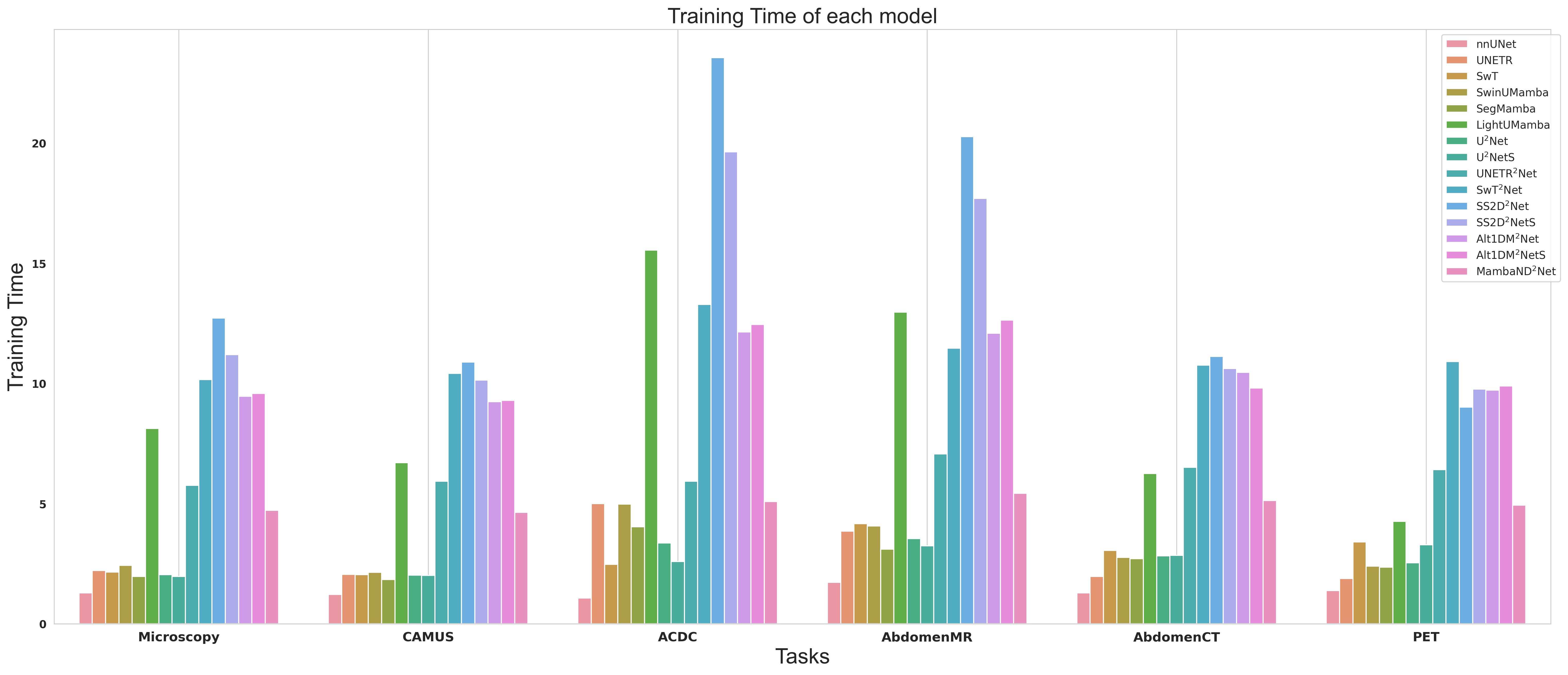}
    \caption{\textit{Time comparison bar plot across different networks and datasets}}
\label{fig:time_barplot}
\end{figure}

\begin{figure}
   \centering
   \includegraphics[width=\textwidth]{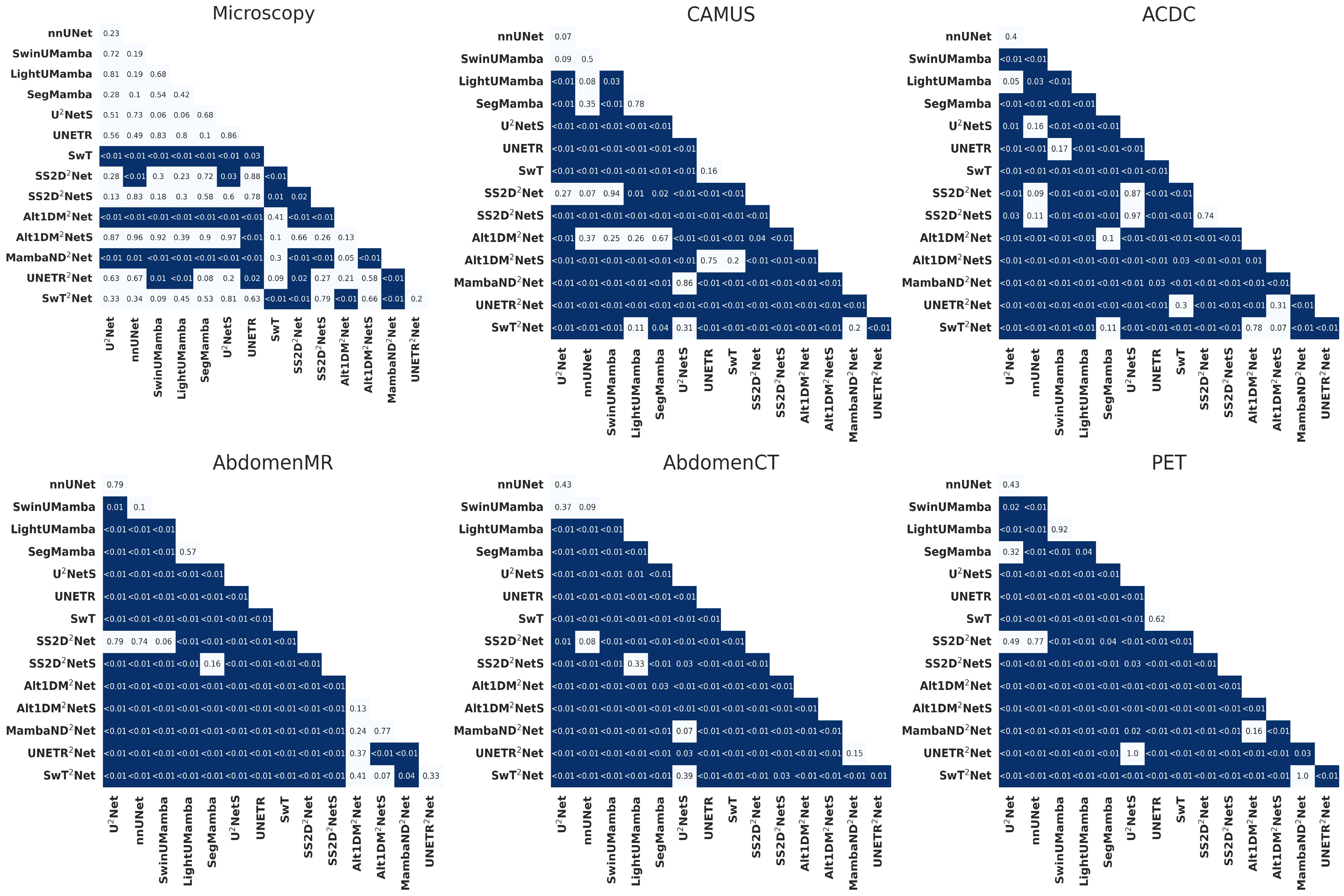}
   \caption{\textit{Comparison of different models' tests}}
   \label{statistics}
\end{figure}
SS2D$^2$Net, a Mamba-based U$^2$Net architecture, performed similarly to nnUNet across all segmentation tasks and imaging modalities. While its performance differences were not statistically significant (Figure~\ref{statistics}) in most datasets, it showed slight improvements over both nnUNet and U$^2$Net. SS2D$^2$NetS, a significantly smaller version of SS2D$^2$Net (about 15 times fewer parameters), performed similarly in microscopy, CAMUS, and ACDC datasets. However, its performance dropped significantly in more complex datasets such as AbdomenMR, AbdomenCT, and PET, indicating limitations in handling higher anatomical variability and structural complexity. Alt1DM$^2$Net, its smaller version, Alt1DM$^2$NetS, and MambaND$^2$Net achieved comparable results in Microscopy, CAMUS, and ACDC datasets. However, their performance was significantly lower in AbdomenMR, AbdomenCT, and PET compared to nnUNet and SS2D$^2$NetS.

\begin{table}
\centering
\caption{Model Parameter Sizes in million across different networks and datasets}
\label{params-table}
\begin{tabular}{|c|c|c|c|c|c|c|}
\toprule
Model Names & Microscopy &  CAMUS &   ACDC & AbdomenMR & AbdomenCT &    PET \\
\midrule
nnUNet                        &       62.2 &   62.2 &   28.7 &      45.4 &      45.4 &   45.5 \\
UNETR                         &      110.7 &  110.3 &  110.3 &     110.4 &     110.3 &  110.3 \\
SwT                           &       39.5 &   39.5 &   39.5 &      39.5 &      39.5 &   39.5 \\
SwinUMamba                    &       26.2 &   26.2 &   26.2 &      26.2 &      26.2 &   26.2 \\
SegMamba                      &       25.3 &   25.3 &   25.3 &      25.3 &      25.3 &   25.3 \\
LightUMamba                   &       5.70 &   5.70 &   5.70 &      5.70 &      5.70 &   5.70 \\
\midrule
U\textsuperscript{2}Net       &       42.0 &   42.0 &   42.0 &      42.1 &      42.1 &   42.3 \\
U\textsuperscript{2}NetS      &       1.10 &   1.10 &   1.10 &      1.10 &      1.10 &   1.20 \\
\midrule
UNETR\textsuperscript{2}Net   &      149.1 &  149.1 &  149.0 &     149.3 &     149.1 &  149.1 \\
SwT\textsuperscript{2}Net     &      172.2 &  172.2 &  172.2 &     172.3 &     172.3 &  172.3 \\
SS2D\textsuperscript{2}Net    &       39.0 &   39.1 &   39.1 &      39.2 &      39.2 &   39.3 \\
SS2D\textsuperscript{2}NetS   &       2.00 &   2.00 &   2.00 &      2.10 &      2.10 &   2.20 \\
Alt1DM\textsuperscript{2}Net    &       8.90 &   8.90 &   8.90 &      8.90 &      8.90 &   8.90 \\
Alt1DM\textsuperscript{2}NetS   &       1.50 &   1.50 &   1.50 &      1.50 &      1.50 &   1.50 \\
MambaND\textsuperscript{2}Net &       39.5 &   39.5 &   39.5 &      39.5 &      39.5 &   39.5 \\
\bottomrule
\end{tabular}
\end{table}

Table~\ref{params-table} presents the number of model parameters (in millions) for each network. The three networks with the highest performance—nnUNet, U$^2$Net, and the proposed SS2D$^2$Net—did not show statistically significant (Figure~\ref{statistics}) differences in performance in most datasets despite variations in the number of parameters. SS2D$^2$Net has 3 million fewer parameters than U$^2$Net and 6.2 to 23.2 million fewer parameters than nnUNet, depending on the task (due to adaptability of nnUnet to the input data). However, in the ACDC dataset, nnUNet had 10.4 million fewer parameters compared to SS2D$^2$Net. This shows SS2D$^2$Net’s efficiency, as it maintains comparable performance while reducing model complexity in most tasks. The number of parameters in Alt1DM$^2$Net, Alt1DM$^2$NetS, and MambaND$^2$Net is significantly and than in SS2D$^2$Net. However, their performance is consistently lower across all tasks, indicating that the reduction in parameters comes at the cost of segmentation accuracy. 

Figure~\ref{fig:time_barplot} presents a comparison of training times across all networks. As shown, nnUNet, which is CNN-based, requires the least training time. In contrast, Mamba-based networks, including SS2D$^2$Net and SS2D$^2$NetS, require significantly longer training times—approximately 5 to 20 times more than nnUNet. This indicates that while Mamba-based architectures achieve comparable performance to nnUNet with fewer parameters, the computational cost of training remains substantially higher.

\newpage
\section{Conclusion and Future Work}

The proposed nnUZoo provides a benchmark for multiple deep learning architectures based on CNNs, Transformers, and Mamba. It includes implementations of different architectures with various training settings, designed to be easy to use with low-code approaches for developing segmentation models across different modalities. CNN models like nnUNet and U$^2$Net demonstrated both speed and accuracy, making them effective choices for medical image segmentation tasks. Transformer-based models, while promising for certain imaging modalities, exhibited high computational costs. Mamba-based X$^2$Net architectures achieved competitive accuracy with no statistically significant difference in most datasets from nnUNet and U$^2$Net, while using fewer parameters. However, they required significantly longer training time, highlighting a trade-off between model efficiency and computational cost. 

While nnUNet itself dynamically adapts to different tasks, the baseline and proposed networks used within its pipeline remain static across different modalities and computational environments. Future work will focus on making these architectures dynamic, allowing them to adapt to imaging modalities and hardware constraints, similar to nnUNet’s planning pipeline. This will enhance the flexibility and efficiency of deep learning-based medical image segmentation. Additionally, self-supervised pre-training, particularly for Transformer and Mamba-based architectures within the nnUNet pipeline, could be explored to potentially enhance model performance.
\newpage

\bibliographystyle{elsarticle-num} 
\bibliography{main.bib}
\end{document}